\documentstyle[12pt,aps,psfig]{revtex}
\newcommand{\de}{\delta}

\newcommand{\eps}{\varepsilon}
\newcommand{\om}{\omega}

\newcommand{\barr}{\begin{array}}
\newcommand{\bea}{\begin{eqnarray}}
\newcommand{\beq}{\begin{equation}}
\newcommand{\ear}{\end{array}}
\newcommand{\eea}{\end{eqnarray}}
\newcommand{\eeq}{\end{equation}}

\newcommand{\bma}{\begin{displaymath}}
\newcommand{\ema}{\end{displaymath}}

\begin{document}
\title{\Large Integral representation of the RPA correlation energy}
\author{\sc F.~D\"onau , D. Almehed and R.G. Nazmitdinov$^{1}$}
\address{\sl Institut f\"ur Kern- und Hadronenphysik, FZ Rossendorf, 
01314 Dresden, Germany\\
$^1$ Bogoliubov Laboratory of Theoretical Physics\\
Joint Institute for Nuclear Research, 141980 Dubna, Russia}
\maketitle

\vspace{0.3cm}

\begin{abstract}
{\sf Using the spectral function F'(z)/F(z) the RPA correlation 
energy and other properties of a finite system can be written as 
a contour integral in a compact way. This yields a transparent 
expression and reduces drastically the numerical efforts 
for obtaining reliable values. The method applied to 
pairing vibrations in rotating nuclei as an illustrative example.}
\end{abstract}

\vspace{0.1cm}


\vspace {0.5cm}

Various problems considered in
many-body physics of finite quantum systems 
can be treated in terms
of a Hamiltonian which is quadratic in the creation and
annihilation operators (see for example \cite{BR}).
The mean field approximation (MF) and the random phase
approximation (RPA) are the well known examples where
the dynamics of complicated system is reducing to dynamics
described by a quadratic Hamiltonian. These approximations
have been proven to be quite successful in many applications
to mesoscopic systems like nuclei \cite{BM,Sol,RS} 
and metallic clusters \cite{Br}.

One of the long standing problems in the RPA is how to
calculate practically the correlation energy of
the Hamiltonian $H=H_0+V$,
where $H_0$ is a mean field Hamiltonian and $V$ is a two-body
residual interaction.
The solution of this problem could allow also to calculate 
properly other physical quantities which depend on the whole 
set of the RPA eigen frequencies.  
The MF description is only a first approximation
and the RPA correlations can essentially improve the  
description of experimental observables.

The Hamiltonian, written in terms of
the RPA phonon operators $O_n^+(O_n)$, takes the simple 
form \cite{BR,RS}
\beq
H=E_{MF}+\sum_{n>0}\om_nO_n^+O_n+\frac{\it P^2}{2\mu}-
\sum_{n>0}\om_n{Y^{n}}^+Y^n
\eeq
Here the first term is the MF energy and 
the second term is the Hamiltonian of the RPA (normal) modes.
The third term is associated with the contribution of a
spurious mode arisen due to a broken continuous
symmetry (for example translation or rotation). The fourth term,
where $Y^n$ is the vector of the backward-going amplitudes 
of the RPA solution, is a correlation energy ${\eps_{\it corr}}$ 
caused by the normal modes.
The eigen frequencies $\om_\nu$ of the RPA  
are derived from the singularities of 
the  RPA response function \cite{PN}
\beq
{\cal R} (\om) = ( 1 - R^0(\om)\chi)^{-1} R^0(\om) \,\,, 
\eeq
where $R^0$ is the known unperturbed response function and
$\chi$ is a coupling constant. Accordingly, these 
frequencies are obtained as the zeros of the determinant
\beq
\label{det}
F(\om) =  det\,(1 - R^0(\om)\chi \,) .    
\eeq
The total correlation energy $E_{\it corr}$, which is essentially the 
energy gain of the RPA ground state relative to the 
MF ground state, can be expressed as
\beq
\label{cren}
E_{\it corr}=-\frac{<\it P^2>_{MF}}{2\mu}-{\eps_{\it corr}} 
\equiv \frac{1}{2}( \sum \om_\nu - \sum E_\mu )-E_{ex} \,\,,
\eeq
where $E_{ex}$ is the known exchange energy 
of the interaction $V$ \cite{BR} and $E_{\mu}$ are 
the poles of the $R^0$. 
The eigen frequencies $\om_\nu$ appear 
in actual situations  as rather closely spaced 
zeros of the determinant $F(\om)$, Eq.(\ref{det}). 
Since the number of roots of $F(\om)$ 
is practically of the order $10^4$, for example,
for heavy nuclei ($A\approx 150$) and 
none of these roots can be neglected,
the calculation of the correlation energy $E_{\it corr}$ 
is known to be rather difficult
even in the case of the separable interactions \cite{Eg}.
In addition, in many physical applications  
it is necessary to study the dependence of the correlation energy 
on the variation of the MF parameters.
The method, proposed in \cite{Sh} for the calculation 
of the rotational dependence of the correlation energy, does not
allow treat the contribution of the spurious modes.
Furthermore, it should be used with a dense grid to approach
the necessary convergence in the integration procedure
to obtain reliable results regarding the contribution of 
the normal modes. The contour integral representation 
presented in our paper remedies these deficiencies
and in addition is a transparent method which has further 
advantages discussed below.

Let us consider a general eigenvalue equation   
\beq
\label{fun}
F(z) = 0
\eeq
where $F(z)$ is a supposed to be an analytical function
in some region of the complex variable $z$. The 
complex continuation of the determinant Eq.(\ref{det}) is an
example for an appropriate function $F(z)$ studied below in detail. 
Another example is any finite matrix eigenvalue problem 
$F(z)= det\,(H-z) = 0$ written in terms of a finite matrix 
$H$, for instance, the Hamiltonian of the
shell model \cite{BG}.
We define for our purpose the spectral function $S(z)$ as
\beq
S(z) = \frac{F'(z)}{F(z)} = \frac{d}{dz}\, log F(z)
\eeq
using the derivative $F'(z)$ in the z-region 
where $F(z)$ is analytical. 
The meaning of $S(z)$ becomes obvious for the example
$F(z)= det\,(H-z)$. Assuming  $H$ is diagonalizable
we may express  $F$ in terms of the eigenvalues
$\om_\nu$ as the product $F(z)=\Pi_\nu (\om_\nu - z)$
and obtain
\beq
S(z) = \frac{F'(z)}{F(z)} = \sum_\nu \frac{1}{\om_\nu - z} 
\eeq  
which yields the spectral decomposition of the
matrix $H$. 
Returning to the general case, it is suggestive to apply 
Cauchy's theorem to the above spectral function $S(z)$.
In particular, we obtain straightforwardly by forming 
the following closed contour integral 
\bea
\label{form}
{\cal I} \,[g(z)]& \equiv& 
\frac{1}{2\pi i} \oint dz\, g(z)\, S(z) = 
\frac{1}{2\pi i}\oint dz \,g(z)\,\frac{F\,'(z)}{F(z)}\\ 
&=&\sum_\nu n_\nu g(\om_\nu) - \sum_\mu m_\mu g(p_\mu)\nonumber   
\eea    
where $g(z)$ is an arbitrary complex function which is analytical 
in the enclosed region. Further, the sum $\nu$ includes all
roots $\om_\nu$ of $F(z)$  and the  sum $\mu$ accounts for
all poles $p_\mu$ of the derivative $F'(z)$
enclosed by the chosen contour. The integers
$n_\nu$ and $m_\mu$ mean the multiplicity of the root $\om_\nu$ and
the order of the pole $p_\nu$, respectively.

The integration formula Eq.(\ref{form}) is now applied to the RPA case.
For the sake of the discussion let us consider a two-body
residual interaction which is a sum of separable multipole-multipole
interactions, i.e.
\bea 
\label{int}
V&=&\sum_{\rho}\chi_{\rho}{\hat Q_{\rho}^+}{\hat Q_{\rho}}\\
{\hat Q_{\rho}}&=&\sum_{\alpha <\beta}q_{\rho}(\alpha \beta)
a_{\alpha}^+a_{\beta}^+ + h.c.={\hat Q_{\rho}^+}\nonumber
\eea
where $a^+$ is a quasiparticle creation operator.
This interaction is widely used in nuclear physics \cite{BM,Sol,RS} and
nowadays it is successfully employed for the description of 
the plasmon frequencies in metallic clusters \cite{Kl}.
The unperturbed matrix response function 
$R^0$  associated with the interaction Eq.(\ref{int}) is defined  
as the matrix
\beq
\label{mel}
R^0_{\rho\sigma}(\om) = \sum_\mu
( \frac{q^*_\rho (\mu) q_\sigma (\mu)}{E_\mu - \om} \,+\,
\frac{q_\rho (\mu) q^*_\sigma (\mu)}{E_\mu + \om} )
\eeq
where the double index $\mu = (i,j)$ is running over all 
independent quasiparticle pairs $i>j = 1,n$ and $E_\mu \equiv e_i + e_j$ 
denotes the two-quasiparticle energy. The function $F(z)$ is 
specified as the determinant Eq.(\ref{det}) with matrix elements 
Eq.(\ref{mel}). The location of its roots $\om_\nu$ is not 
needed but these RPA roots are known to lie on the real axis 
symmetrically around the origin and in addition appear at $z=0$, 
if there are spurious RPA solutions \footnote{the case of complex
RPA solutions is normally excluded but could be 
similarly treated}. One can easily obtain the derivative 
$F'(z)$ to form $S(z)$. Apparently, the poles of  $F'(z)$ coincide
with the known two-quasiparticle energies $E_\mu$.  
For our goal the appropriate contour of the integration 
Eq.(\ref{form}) goes around all positive roots of $F(z)$ and 
likewise all two-quasiparticle poles as  shown in Fig.1.
Hence the desired contribution to the RPA correlation 
energy Eq.(\ref{cren}) is obtained as the contour integral  
\beq
\label{fin}
\frac{1}{2}( \sum \om_\nu - \sum E_\mu )\equiv 
\frac{1}{4\pi i}\oint dz \,z \,\frac{F\,'(z)}{F(z)}
\eeq
Usually, the  integration will be done numerically. The 
crucial practical  advantage of the formula Eq.(\ref{fin}) is that  
we are free to choose the rectangular contour (Fig.1) 
sufficiently distant from the poles 
such that the spectral function 
$S(z) = F'(z)/F(z)$  becomes smooth
while integrating on this path. 

\begin{figure}[htb]
\centerline{\psfig{figure=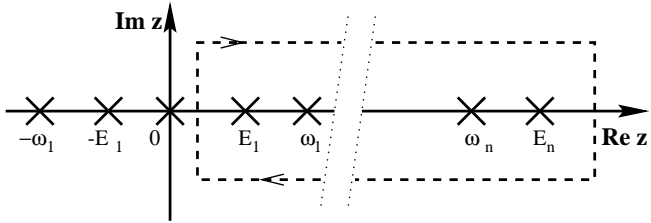,angle=0,height=3cm}}
\vspace{2 mm}
\caption{A schematic picture of the integration contour (dashed line) 
in the complex plane. The roots $\om_{\nu}$ and the poles $E_{\mu}$ 
of the function $F(z)$ are marked with crosses.}
\label{fig1}
\end{figure}

Then, the necessary grid needs not to be dense any 
more. In practical cases considered so far the number of
integration points were reduced by a factor $10^2$
without loss of a precision which reduces the CPU time 
drastically. The contribution of the possible spurious 
RPA solutions with $\om_0=0$ is correctly accounted for, since 
the equivalent sum expression Eq.(\ref{cren}) 
implies such contributions. 
Notice that, when integrating in Eq.(\ref{fin}) separately
around such a spurious pole at $z=0$, there arises
no a finite energy contribution due to the $z-$factor
in the integrand. The precision of the         
numerical integration of the correlation energy $E_{corr}$
can be conveniently checked by evaluating the analogous contour 
integral for $g(z) = 1$ which gives according Eq.(\ref{form}) 
an integer number.
For the RPA spectral function $S(z)$ the difference between
the number of RPA roots and the number of quasiparticle states
is counted, i.e. one expects a zero if there are no spurious 
$\om_0=0$ solutions. 

We specify discussion to the RPA spectral function $S(z)$
derived from the determinant Eqs.(\ref{det}),(\ref{mel}). 
However,  a similar treatment may hold for more general cases. 
Even, if the characteristic  function $F(z)$, Eq.(\ref{fun}), can be  
defined only numerically in the vicinity of an appropriate 
integration path one may calculate the spectral 
function $S(z)$ and apply the integration method.  
Now, we are going to employ the free choice of any physically
relevant weight function $g(z)$ in the integral formula Eq.(\ref{form}).
In addition, also the integration contour 
can be adapted to the physical problem under study.
For  $g(z) \equiv 1$ information on the number and distribution 
of levels can be obtained. First, we define the two-quasiparticle
spectral function by 
\beq
S^\circ (z) = \sum_\mu \frac{1}{E_\mu - z} \,.
\eeq 
Let us assume for simplicity, that the derivative 
$F\,'(z)$ of the RPA determinant Eqs.(\ref{det}), (\ref{mel}), 
has simple poles, i.e. $m_\nu = 1$. Then, in the spectral function
\beq
\tilde{S}(z) = S(z) - S^\circ(z) 
\eeq 
the two-quasiparticle poles are removed. 
Hence the integral ${\cal I}[g=1]$ using the spectral function 
$\tilde S$ is counting the number of all RPA states $\om_\nu$
enclosed in the contour. The presence of a spurious solution
at $z=0$ can be checked when calculating  ${\cal I}[g=1]$ with a narrow
rectangular contour around zero which cannot be seen directly
from the correlation energy Eq.(\ref{fin}).
The simplest form of a level density distribution 
can be calculated analogously by slices like ${\cal I}[g=1]/\Delta\om $
by integrating repeatedly along rectangular stripe contours 
around a given $\om-$value with an appropriate width $\Delta\om$.
A more refined representation of the level distribution 
follows from Eq.(\ref{form}) when inserting the Lorentzian weight function
\beq
g(z) \equiv L(z) = \frac{\Delta^2}{(\om - z)^2 + \Delta^2}
\eeq
which has poles at $z=\om \pm i\Delta $. 
The RPA spectral function $\tilde S(z)$
yields the relation
\beq
\label{den}
D(\om,\Delta) \equiv\,
 \sum_\nu \frac{\Delta^2}{(\om - \om_\nu)^2 + \Delta^2} \,=\,
\Delta Im\tilde S(\om-i\Delta)\,  + \,{\cal I}[L(z)]\,       
\eeq 
where the sum includes all eigenvalues $\om_\nu$ within the chosen 
contour. 
Actually it is the  r.h. side of this relation 
which can be evaluated numerically and thus it provides us the 
RPA level distribution $D(\om,\Delta)$ averaged by the width parameter
$\Delta$. Taking the contour at infinity outside all poles
(assuming a finite number of roots) the integral becomes zero
due to the Lorentzian fall off. Hence the integral term 
in Eq.(\ref{den}) is the Lorentzian ''background'' contribution from the
roots not included in the contour.  
Again, the contour integral involved can be calculated 
conveniently with a smooth integrand as mentioned above.

To demonstrate the viability and utility of the method, 
we have applied it to analyze the influence of the correlation 
energy on the description of the g-band of $^{178}W$. 
For our purpose the tilted axis cranking model 
(TAC) \cite{Fr} represents the proper  Hartree-Fock-Bogoliubov 
(HFB) theory which permits the 
calculations of the energies and intra band probabilities. 
As a residual interaction we consider the standard
pairing field $V=-GP^+P$. It is well known that the HFB 
theory does not provide an adequate description 
of the transition region where the pair field strongly fluctuates.
Therefore we use the version of the TAC that includes the particle 
number projection (PNP) and improves the description of the pair
correlations in high-K bands \cite{Al1}. Fig.2 shows the results 
for different Routhian energies of the g-band calculated with the 
TAC alone, with the TAC which incorporates the PNP, and
the TAC results which include the RPA correlation energy due to pairing
vibrations, Eq.(\ref{cren}). On the whole region of a rotational 
frequency the RPA correlation energy lowers substantially the TAC 
Routhian energy and, consequently, improves the description of the 
g-band in comparison with the TAC combined with the PNP. 

\begin{figure}[htb]
\centerline{\psfig{figure=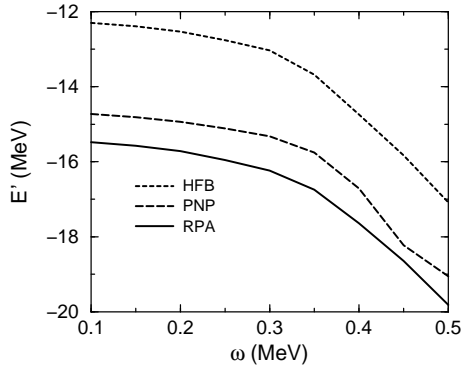,angle=-90,height=5cm}}
\vspace{2 mm}
\caption{The g-band of $^{178}W$. Different Routhian energies are presented:
for the TAC (short-dashed line), for the TAC which includes
the PNP (long-dashed line) and for the TAC with the 
RPA correlation energy (solid line).}
\label{fig2}
\end{figure}

Using the same method, we can evaluate the contribution of
the RPA correlations to any physical quantity in the 
correlated ground state $\de Q_{RPA}=<Q>_{RPA}-<Q>_{MF}$ where
$\de Q_{RPA}=\sum_{n>0}\sum_{ijl}q_{ij}Y^{n*}_{il}Y^n_{jl}$ \cite{Eg,Naz}.
For example, the aligned angular momentum calculated in the 
cranking HFB theory is larger than the experimental alignment. 
The consideration of the additional term arisen due to 
the RPA correlations leads to better agreement with experimental 
data. Notice that the strength function method \cite{KN} 
may be considered as an alternative approach, however, it would 
treat the spurious contributions and $\varepsilon_{\it corr}$ 
separately. The comprehensive analysis of results obtained with 
different approaches will be discussed in the forthcoming paper \cite{Al}.

In conclusion, the transparent practical method has been 
developed which drastically reduces the numerical
efforts to calculate the integral characteristics renormalized 
by residual interactions in finite quantum systems. 
In particular, the correlation energy 
which arises in the RPA can be estimated with the necessary
accuracy. In the considered example of $^{178}W$ the method 
demonstrates a remarkable improvement of the description of the 
g-band in comparison with the particle number projection approach.

\vskip 1cm
 
We are grateful to  S.Frauendorf for useful discussions.
This work was supported in part by the Heisenberg-Landau program
of the JINR.

\end{document}